\documentclass[conference]{IEEEtran}
\usepackage{amssymb}
\usepackage[cmex10]{amsmath}
\usepackage{theorem}
\usepackage{latexsym}
\usepackage{graphicx}
\IEEEoverridecommandlockouts
\theoremheaderfont{\bf}
\theorembodyfont{\sl}

{\theorembodyfont{\rm} }
{\theorembodyfont{\rm} }
{\theorembodyfont{\rm} }
{\theorembodyfont{\rm} }

{\theorembodyfont{\rm} }

{\theorembodyfont{\rm}}
{\theorembodyfont{\rm}}

\newcommand{\ie}{{\em i.e., }}
\newcommand{\eg}{{\em e.g., }}
\newcommand{\cf}{\emph{cf.\ }}

\newcommand{\F}{{\mathbb F}}

\newcommand{\CC}{{\mathcal C}}

\newcommand{\PP}{{\mathcal P}}

\newcommand{\SSS}{{\mathcal S}}

\newcommand{\cI}{{\mathcal I}}
\newcommand{\A}{{\mathcal A}}

\newcommand{\cT}{{\mathcal T}}
\newcommand{\cF}{{\mathcal F}}

\newcommand{\cJ}{{\mathcal J}}

\newcommand{\Bf}{{\mathfrak B}}

\newcommand{\Bfu}{{\mathfrak{B}_\mathrm{u}}}

\newcommand{\zerob}{{\mathbf 0}}
\newcommand{\oneb}{{\mathbf 1}}
\newcommand{\ab}{{\mathbf a}}
\newcommand{\gb}{{\mathbf g}}
\renewcommand{\sb}{{\mathbf s}}
\newcommand{\vb}{{\mathbf v}}

\newcounter{alp}
\newcounter{ara}
\newcounter{rom}

\newcommand{\openbox}{\leavevmode
     \hbox to.77778em{%
     \hfil\vrule
     \vbox to.675em{\hrule width.6em\vfil\hrule}%
     \vrule\hfil}}
\newcommand{\qed}{\hspace*{1cm}\hspace*{\fill}\openbox}

\begin{document}
%
\title{Observability, Controllability and Local Reducibility \\ of Linear Codes on Graphs}

\author{
\IEEEauthorblockN{G.~David Forney, Jr.}
\IEEEauthorblockA{Laboratory for Information and Decision Systems\\
Massachusetts Institute of Technology\\
Cambridge, MA 02139\\
Email: forneyd@comcast.net
}
\and
\IEEEauthorblockN{Heide Gluesing-Luerssen}
\IEEEauthorblockA{University of Kentucky\\
Department of Mathematics\\
Lexington, KY 40506\\
Email: heide.gl@uky.edu
}
\thanks{The work of the second author was supported in part by National Science
Foundation grant \#DMS-0908379.}
\thanks{This conference paper is based on the preprint~\cite{FGL12}.}
}

\maketitle

\begin{abstract}
This paper is concerned with the local reducibility properties of linear realizations of codes on finite graphs.

Trimness and properness are dual properties of constraint codes.  A linear realization is locally reducible if any constraint code is not both trim and proper.  On a finite cycle-free graph, a linear realization is minimal if and only if every constraint code is both trim and proper.    

A linear realization is called observable if it is one-to-one, and controllable if all constraints are independent.  Observability and controllability are dual properties.  An unobservable or uncontrollable realization is locally reducible. A parity-check realization is uncontrollable if and only if it has redundant parity checks.  A tail-biting trellis realization is uncontrollable if and only if its trajectories partition into disconnected subrealizations.  General graphical realizations do not share this property.  

\end{abstract}


\IEEEpeerreviewmaketitle

\section{Introduction}\label{S-Intro}
For conventional linear state-space (trellis) realizations of codes or systems, minimality is well understood and can be characterized in many ways; see \cite{V98, F11} and the references therein.

For realizations on more general graphs, the situation is much less studied and more complicated.
In particular, a given code does not have a unique minimal realization on a graph with cycles;
see \cite{KV03} for a detailed discussion of the specific class of tail-biting trellis realizations.
Yet it has long been known that realizations on graphs with cycles can be much simpler, and therefore may allow more powerful iterative decoding algorithms.

In this paper, we will study properties of linear realizations on general graphs that allow ``local reduction" operations.  For instance, we will show that local reduction is possible if any constraint code in the realization is not both trim and proper;  moreover, when the graph of the realization is cycle-free, the converse is  sufficient for minimality.  We go on to show that a realization is locally reducible if it is not observable (one-to-one) or uncontrollable (has dependent constraints).  We show that the effects of uncontrollability are similar for tail-biting trellis realizations to those for conventional trellis realizations, but that this result does not generalize straightforwardly to the general case.

\section{Notation and terminology}\label{S-NT}
All of our realizations will be finite and linear.  We will generally use the notation and terminology of \cite{F01} and \cite{KV03}.

A linear code $\CC$ over a finite field $\F$ is a subspace of a \emph{symbol sequence space} $\A = \Pi_{k \in \cI_\A} \A_k$, where each
\emph{symbol alphabet} $\A_k$ is a finite-dimensional vector space over $\F$, and $\cI_\A$ is a finite index set.
For a \emph{state realization} of $\CC$, we define also a set $\{\SSS_j: j \in \cI_\SSS\}$ of \emph{state spaces} $\SSS_j$ and a set
$\{\CC_i: i \in \cI_\CC\}$  of local \emph{constraint codes} $\CC_i$, where each
constraint code $\CC_i$ involves some subsets
$\A^{(i)} = \prod_{k \in \cI_{\A^{(i)}}} \A_k$ and $\SSS^{(i)} = \prod_{j \in \cI_{\SSS^{(i)}}} \SSS_j$ of the symbol and state
variables, respectively.
In a linear realization, each state space $\SSS_j$ and each constraint code $\CC_i$ is a vector space over $\F$. 
The \emph{state sequence space} is $\SSS=\prod_{j \in \cI_{\SSS}} \SSS_j$.

The \emph{behavior} of a realization is the set $\Bf$ of all pairs  $(\ab, \sb) \in \A \times \SSS$ such that all constraints are
satisfied; \ie $(\ab^{(i)}, \sb^{(i)}) \in \CC_i$ for all $i$.
The code $\CC$ that it \emph{generates} or \emph{realizes} is the set of all symbol sequences $\ab \in \A$ that appear in some
$(\ab, \sb) \in \Bf$;  \ie the projection $\Bf_{|\A}$ of $\Bf$ on $\A$.

A state realization is called \emph{normal} if every symbol variable is involved in precisely one constraint code, and every state
variable is involved in precisely two constraint codes.
As shown in \cite{F01}, any realization may be straightforwardly ``normalized" by introducing replica variables and equality
constraints.
Therefore we will assume henceforth that all realizations are normal.
A normal realization has a natural graphical representation, a \emph{normal graph}, in which constraint codes are represented
by vertices, states by ordinary edges, and symbol variables by half-edges; see \cite{F01}.
We will assume that all normal graphs are connected.


A \emph{trellis realization} is a normal realization in which every constraint code involves precisely two state variables.
Thus the graph of a trellis realization must be either
a finite chain graph, called a \emph{conventional} trellis realization, or
a single-cycle graph, called a \emph{tail-biting} trellis realization.
We will depict trellis realizations by traditional \emph{trellis diagrams}, in which all branches, states and symbols are shown
explicitly.

A realization of a code~$\CC$ will be called \emph{observable} if for each $\ab \in \CC$ there is precisely one pair $(\ab, \sb) \in \Bf$.
We define the \emph{unobservable behavior} $\Bfu$ as the set of pairs $(\zerob, \sb) \in \Bf$;
thus a realization is observable iff $\Bfu$ is trivial.

A realization will be called \emph{state-trim} if each state appears on a valid trajectory;  \ie if the projection of~$\Bf$ onto
each~$\SSS_j$ is $\SSS_j$ (surjective).
Similarly, using trellis terminology,  a realization will be called \emph{branch-trim} if the projection of~$\Bf$ onto
each constraint code~$\CC_i$ is surjective.
A  state-trim and branch-trim  realization is called \emph{reduced}  \cite{KV03}.

A constraint code~$\CC_i$ will be called \emph{trim} if the projection of~$\CC_i$ onto every state space~$\SSS_j$ that is involved in~$\CC_i$ is surjective,
and  \emph{proper} if it has
no nonzero codewords whose support is a single state space $\SSS_j$.

As in~\cite{KV03}, a realization with state spaces~$\SSS_j$ is called \emph{minimal} if there exists no 
realization of the same code with state spaces~$\tilde{\SSS}_j$ that has the same graph topology (\ie the same index sets
$\cI_\A, \, \cI_\SSS,\,\cI_\CC,\,\cI_{\A^{(i)}},\,\cI_{\SSS^{(i)}}$) such that $\dim\tilde{\SSS}_j\leq\dim\SSS_j$
for all~$j$, with at least one strict inequality.  It is well known that on a cycle-free graph a minimal realization is unique up to isomorphism;  however,  on a graph with cycles, there is in general no unique minimal realization. 

In this paper, a realization will be called \emph{locally reducible} if there is a replacement of one state space $\SSS_j$ by a smaller state space, and a corresponding reduction of the constraint codes involving $\SSS_j$, such that the resulting realization realizes the same code.  (More general notions of local reducibility will be considered in \cite{GLF12}.)  We use two dual methods of reducing a state space, namely restricting and taking quotients;  these operations will be called \emph{trimming} and \emph{merging}, respectively.

\subsection{Duality}\label{S-Duality}

We briefly recall some of the basic duality principles that will be used heavily in this paper, following \cite{F01, FT04}.

If $V$ is a finite-dimensional vector space over a finite field $\F$, then its linear-algebra \emph{dual space}  $\hat{V}$ is a vector space over $\F$ of the same dimension such that there is a well-defined inner product $\langle \cdot , \cdot \rangle: V \times \hat{V} \to \F$.
Given any basis of~$V$,  there exists a \emph{dual basis} for~$\hat{V}$ such that the inner product of two vectors in~$V$ and~$\hat{V}$ is equal to the dot product of their respective coordinate vectors.  Two elements $v \in V, \hat{v} \in \hat{V}$ are \emph{orthogonal} if $\langle v, \hat{v} \rangle = 0$.

If~$W\subseteq V$ is a subspace, then its \emph{orthogonal space} $W^{\perp}\subseteq\hat{V}$ is  the space of all vectors in~$\hat{V}$ that are orthogonal to all vectors in~$W$. We have $(W^{\perp})^{\perp}=W$ and $\hat{W}\cong \hat{V}/W^\perp$, implying $\dim W + \dim W^\perp = \dim \hat{V} = \dim V$.

If $V = \Pi_k V_k$ is a finite direct product of vector spaces $V_k$, then 
$\hat{V} = \Pi_k \hat{V}_k$, with inner product given by the componentwise sum $\langle{\vb},{\hat{\vb}}\rangle = \sum_k \langle{v_k},{\hat{v}_k}\rangle$.
If $W = \Pi_k W_k$ is a direct product of subspaces $W_k \subseteq V_k$, then $W^\perp = \Pi_k W_k^\perp \in \hat{V}$.

A key duality lemma for linear codes is \emph{projection/cross-section duality} \cite{F01}.  For any subset $\cJ \subseteq \cI_\A$ of an index set $\cI_\A$, the projection map $P_\cJ: \A \to \A^\cJ = \prod_{k\in\cJ} \A_k$ is defined by the map $\ab = \{a_k: k \in \cI_\A\} \mapsto \ab^\cJ = \{a_k: k \in \cJ\}$.  The \emph{projection} $\CC_{|\cJ}$ of a linear code $\CC$ defined on $\cI_\A$ on $\cJ$ is then defined as the image of $P_\cJ: \CC \to \A^\cJ$, a subspace of $\A^\cJ$.  The \emph{cross-section}  of $\CC$ on $\cJ$ is defined as $\CC_{:\cJ} = \{\ab^\cJ \in \CC^\cJ \mid (\ab^\cJ, \zerob^{\cI_\A-\cJ}) \in \CC\}$, a subspace of $\CC_{|\cJ}$.  Then we have:

\pagebreak
\vspace{1ex}
\noindent
\textbf{Projection/cross-section duality}.  If $\CC$ and $\CC^\perp$ are orthogonal linear codes defined on $\cI_\A$, and $\cJ \subseteq \cI_\A$, then $\CC_{:\cJ}$ and $(\CC^\perp)_{|\cJ}$ are orthogonal linear  codes defined on $\cJ$.  \vspace{1ex}

In this paper, our main tool will be \emph{normal realization duality}, which follows directly from projection/cross-section duality \cite{F01}.  Given a normal linear realization, its \emph{dual realization} is defined as the normal linear realization with the same graph topology in which the
variable alphabets~$\A_k,\,\SSS_j$ are replaced by their dual spaces $\hat{\A}_k,\,\hat{\SSS}_j$, the constraint codes~$\CC_i$ are 
replaced by their orthogonal codes $\CC_i^{\perp}$, and the
sign of each dual state variable is inverted in one of the two constraints in which it is involved.  Then we have:

\vspace{1ex}
\noindent
\textbf{Normal realization duality}.  If a normal realization realizes a linear code~$\CC$, then its dual realization realizes the orthogonal linear code~$\CC^\perp$.  \vspace{1ex}

\section{Trimness and properness}\label{S-TP}

To repeat, a linear constraint code $\CC_i$ is trim if the projection of $\CC_i$ onto every state space $\SSS_j$ that is involved in $\CC_i$ is $\SSS_j$, and  proper if there are no nonzero codewords of $\CC_i$ whose support is a single state space $\SSS_j$.

As observed in \cite{GLW11b}, these are dual properties:

\vspace{1ex}
\noindent
\textbf{Theorem 1}. A linear constraint code $\CC_i$ is trim if and only if the orthogonal constraint code $\CC_i^\perp$ is proper.

\vspace{1ex}
\noindent
\textit{Proof}:  $\CC_i$ is not trim if the projection  $(\CC_i)_{|\SSS_j}$ of $\CC_i$ on any state space $\SSS_j$ involved in $\CC_i$ is a proper subspace of $\SSS_j$.  $\CC_i^\perp$ is not proper if the cross-section $(\CC_i^\perp)_{:\hat{\SSS}_j}$  is nontrivial.  By projection/cross-section duality, $(\CC_i^\perp)_{:\hat{\SSS}_j}$ is the orthogonal code to the projection $(\CC_i)_{|\SSS_j}$.  Thus $(\CC_i^\perp)_{:\hat{\SSS}_j}$ is nontrivial if and only if $(\CC_i)_{|\SSS_j}$ is a proper subspace of $\SSS_j$.  \qed \vspace{1ex}  

This leads immediately to conditions for local reducibility:

\vspace{1ex}
\noindent
\textbf{Theorem 2}. A realization containing a constraint code $\CC_i$ that is not both trim and proper is locally reducible.

\vspace{1ex}
\noindent
\textit{Proof}:  A constraint code $\CC_i$ that is not trim may obviously be locally reduced by restricting $\SSS_j$ to $\SSS_j' = (\CC_i)_{|\SSS_j}$, without changing the code $\CC$ realized by the realization.  

Dually, a linear constraint code $\CC_i$ that is not proper (\ie involves a state space $\SSS_j$ such that $\cT_j = (\CC_i)_{:\SSS_j}$ is nontrivial) may be locally reduced by mapping $\SSS_j \to \SSS_j/\cT_j$ via the natural map $s_j \mapsto s_j + \cT_j$, with corresponding maps in all constraint codes that involve $\SSS_j$, without changing the code $\CC$ realized by the realization.   \qed \vspace{1ex}

Since taking quotients as above preserves linearity, it is the appropriate notion of ``merging" for linear realizations.  

Finally, we have the following remarkably simple result:

\vspace{1ex}
\noindent
\textbf{Theorem 3 (trim + proper = minimal)}. For a linear realization on a finite cycle-free graph, the following are equivalent:
\begin{enumerate}
\item The realization is minimal.
\item Every constraint code is both trim and proper. 
\item The state space associated with any edge $\SSS_j$ is isomorphic to $\CC_{|\PP_j}/\CC_{:\PP_j}$, where $\PP_j$ is the ``past" of $\SSS_j$.
\end{enumerate}

\vspace{1ex}
\noindent
\textit{Sketch of proof} (see \cite{FGL12}): \\ ($1 \Rightarrow 2$)  By Theorem 2, if any constraint code is not both trim and proper, then the realization is locally reducible.   \\
($2 \Rightarrow 3$)
 In a cycle-free graph, every edge $\SSS_j$ is a cut set, whose removal partitions the graph into two disconnected subgraphs, $\PP_j$ and $\cF_j$.  If the graph is finite, then each edge has a finite past depth $d_j$ (maximum distance to any leaf in $\PP_j$). 
 By recursion on $d_j$, we can show that trimness implies that every state $s_j \in \SSS_j$ is reached by some symbol configuration $\ab^\PP \in \CC_{|\PP_j}$, and properness implies that every $\ab^\PP \in \CC_{|\PP_j}$ reaches a unique state $s_j \in \SSS_j$.  This implies that the set of sequences in $\PP_j$ that reach the zero state in $\SSS_j$ is precisely the cross-section $\CC_{:\PP_j}$.  Moreover, by linearity, if $\ab^\PP(s_j) \in \CC_{|\PP_j}$ is any sequence that reaches another state $s_j \in \SSS_j$, then the set of all sequences that reach that state is $\CC_{:\PP_j} + \ab^\PP(s_j)$, a coset of $\CC_{:\PP_j}$ in $\CC_{|\PP_j}$.  Thus the realization induces a one-to-one map between $\SSS_j$ and the quotient space $\CC_{|\PP_j}/\CC_{:\PP_j}$, which by linearity is an isomorphism. \\
($3 \Rightarrow 1$) Since $\SSS_j \cong \CC_{|\PP_j}/\CC_{:\PP_j}$, and by the same argument $\SSS_j \cong \CC_{|\cF_j}/\CC_{:\cF_j}$, we have that $\CC$ is the union of the cosets $\{\CC_{:\PP_j} \times \CC_{:\cF_j} + (\ab^\PP(s_j),  \ab^\cF(s_j)): s_j \in \SSS_j\}$, where $\{\ab^\PP(s_j): s_j \in \SSS_j\}$ and $\{\ab^\cF(s_j): s_j \in \SSS_j\}$ are sets of coset representatives for $\CC_{|\PP}/\CC_{:\PP}$ and $\CC_{|\cF}/\CC_{:\cF}$, respectively. This implies that no further merging of states is possible, so the realization is minimal.    \qed \vspace{1ex}

Although the result "trim + proper = minimal" has long been known for conventional trellis realizations \cite{V98}, it seems to be new for realizations on more general cycle-free graphs.  

The proof of Theorem 3 shows constructively that a finite linear cycle-free realization is minimal if and only if every state space $\SSS_j$ is isomorphic to $\CC_{|\PP_j}/\CC_{:\PP_j} \cong \CC_{|\cF_j}/\CC_{:\cF_j}$which is the essence of the State Space Theorem \cite{F01}.

This result immediately suggests straightforward iterative minimization algorithms, involving a finite series of local reductions.

Finally, we observe that the ``shortest basis" approach to minimality that is used for conventional linear trellis realizations (see \cite{F11} and references therein) cannot be extended to general cycle-free realizations, because it relies on the ``product construction" for trellis realizations,  which generally does not exist for more general cycle-free graphs.  
In this respect, the ``trim + proper" approach to minimality may be regarded as more basic than the ``shortest basis" approach.

\section{Observability and controllability}\label{S-OC}

The behavior  $\Bf \subseteq \A \times \SSS$ of a linear realization is defined by a system of $\sum_i \dim \CC_i^\perp$ linear homogeneous constraint equations.  The constraint system is \emph{independent} if and only if the dimension of the solution space $\Bf$ is equal to the number of variables minus the number of equations;  \ie iff 
$
\dim \Bf = \dim \A + \dim \SSS - \sum_i \dim \CC_i^\perp.
$

We now show that a linear realization has independent constraints if and only if its dual realization is observable. Because of the classical duality between observability and controllability, we will subsequently call a realization with independent constraints \emph{controllable}.

\vspace{1ex}
\noindent
\textbf{Theorem 4}. A normal linear realization is controllable (has independent constraints) if and only if its dual realization is observable.  Moreover, it is controllable if and only if 
$
\dim \Bf =  \sum_i \dim \CC_i - \dim \SSS. 
$

\vspace{1ex}
\noindent
\textit{Proof}:  In view of the normal degree restrictions, the only way that a nontrivial sum of constraints $(\hat{\ab}^{(i)}, \hat{\sb}^{(i)}) \in \CC_i^\perp$ can equal zero is if all symbol values $\hat{a}_k$ equal zero, and for each pair of dual state values $\hat{s}_j, \hat{s}_j' \in \hat{\SSS}_j$, we have $\hat{s}_j = -\hat{s}_j'$.  But these are precisely the conditions such that $(\hat{\zerob}, \hat{\sb})$ is a valid nonzero trajectory in the dual realization;  \ie for the dual realization to be unobservable.  

For the formula, we start with $\dim \Bf = \dim \A + \dim \SSS - \sum_i \dim \CC_i^\perp$, substitute  $\dim \CC_i^\perp = \dim \A^{(i)} + \dim \SSS^{(i)} - \dim \CC_i$, and note that  $\sum_i \dim \A^{(i)} = \dim \A$ and $\sum_i \dim \SSS^{(i)} = 2 \dim \SSS$, by the normal degree restrictions.
\qed \vspace{1ex}

We use the term ``controllable" even though: (a) this term was introduced at a time when linear system theory was embedded in control theory, which is not our context here;  (b) although the behavior of uncontrollable tail-biting trellis realizations is similar to that of uncontrollable conventional trellis realizations, as we will see below in Theorem 7, such uncontrollability properties do not necessarily extend to realizations on general graphs, as we will see in Example 2.  The reader who is not so interested in continuity with classical linear system theory might therefore prefer terms like ``one-to-one" and ``independent" to ``observable" and ``controllable."

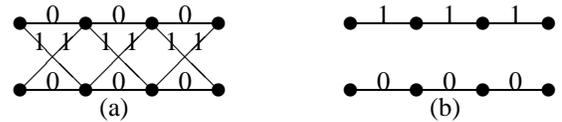
\begin{figure}[h]
\setlength{\unitlength}{5pt}
\centering
\begin{picture}(45,8)(-2, 9)
\put(25,10){\circle*{1}}
\put(25,15){\circle*{1}}
\put(25,10){\line(1,0){5}}
\put(25,15){\line(1,0){5}}
\put(27,10){0}
\put(27,15){1}
\put(30,10){\circle*{1}}
\put(30,15){\circle*{1}}
\put(30,15){\line(1,0){5}}
\put(30,10){\line(1,0){5}}
\put(32,10){0}
\put(32,15){1}
\put(35,10){\circle*{1}}
\put(35,15){\circle*{1}}
\put(35,15){\line(1,0){5}}
\put(35,10){\line(1,0){5}}
\put(37,10){0}
\put(37,15){1}
\put(40,10){\circle*{1}}
\put(40,15){\circle*{1}}
\put(6,8){(a)}
\put(0,10){\circle*{1}}
\put(0,15){\circle*{1}}
\put(0,10){\line(1,0){5}}
\put(0,15){\line(1,0){5}}
\put(0,10){\line(1,1){5}}
\put(0,15){\line(1,-1){5}}
\put(2,10){0}
\put(2,15){0}
\put(1,13){1}
\put(3,13){1}
\put(5,10){\circle*{1}}
\put(5,15){\circle*{1}}
\put(5,15){\line(1,0){5}}
\put(5,10){\line(1,0){5}}
\put(5,15){\line(1,-1){5}}
\put(5,10){\line(1,1){5}}
\put(7,10){0}
\put(7,15){0}
\put(6,13){1}
\put(8,13){1}
\put(10,10){\circle*{1}}
\put(10,15){\circle*{1}}
\put(10,10){\line(1,0){5}}
\put(10,15){\line(1,0){5}}
\put(10,10){\line(1,1){5}}
\put(10,15){\line(1,-1){5}}
\put(12,10){0}
\put(12,15){0}
\put(11,13){1}
\put(13,13){1}
\put(15,10){\circle*{1}}
\put(15,15){\circle*{1}}
\put(31,8){(b)}
\end{picture}
\caption{Dual unobservable and uncontrollable tail-biting trellis realizations.}
\label{Fig2a}
\end{figure}

\vspace{1ex}
\noindent
\textbf{Example 1} (\cf \cite[Fig.\ 5]{KV03}).  The binary linear $(3,2,2)$ block code $\CC = \{000, 110, 101, 011\}$ may be realized by the linear tail-biting trellis realization shown in Fig.\ \ref{Fig2a}(a), with three binary symbol alphabets, three binary state spaces $\SSS_0 =  \SSS_1 = \SSS_2 = \{0,1\}$, and three constraint codes $\CC_0 = \CC_1 = \CC_2 = \{000, 110, 101, 011\}$, where $\CC_2$ involves $\SSS_2$ and $\SSS_0$.  This is a product realization generated by $\langle \underline{11}0, 0\underline{11}, \underline{1}0\underline{1} \rangle$, with the indicated circular spans.
Because the all-zero symbol sequence is realized by two trajectories, this realization is unobservable.   

%

The dual linear tail-biting trellis realization is shown in Fig.\ \ref{Fig2a}(b). (No sign inverters are needed because the field is $\F_2$.)  The dual realization realizes the orthogonal $(3,1,3)$ code $\CC^\perp = \{000, 111\}$;  however, by Theorem 4 it is uncontrollable, since it is the dual to an unobservable realization.   Explicitly, the three constraints  corresponding to the branches $101$ in the three primal constraint codes (\ie the components of the nonzero trajectory $(\zerob, \oneb)$ in the primal behavior) are dependent.  

We shall see in Theorem 7 that the fact that the dual behavior $\Bf^\circ$ comprises two disjoint paths is another proof that the dual realization is uncontrollable. Alternatively, it is uncontrollable because $\dim \Bf^\circ = 1$, $\sum_i \dim \CC_i^\perp = 3$ and $\dim \hat{\SSS} = \sum_i \dim \hat{\SSS}_i = 3$.
  \qed  \vspace{1ex}
  
  \subsection{Generator and parity-check realizations}
  
  We next consider the familiar classes of generator and parity-check realizations,  and determine their observability and controllability properties.
  
  A \emph{generator realization} of a linear code $\CC \subseteq \F^n$ is specified by a set of $\ell$ generator $n$-tuples $\gb_i \in \F^n$ such that $\CC$ is the set of all linear combinations $\ab = \sum_i \alpha_i \gb_i$.  The generators are \emph{linearly independent} if  $\dim \CC = \ell$.  The realization has $n$ symbol variables $A_k$,  $\ell$ equality constraints that generate up to $n$ replicas $\alpha_{ik} = \alpha_i$ of each of the coefficients $\alpha_i$, and $n$ linear constraint codes that enforce the constraints $a_k = \sum_i \alpha_{ik} g_{ik}$.  (If $g_{ik} = 0$, then the replica $\alpha_{ik}$ may be omitted.)
  
A \emph{parity-check realization} of a linear code $\CC$ is the dual to a generator realization of $\CC^\perp$.  Parity-check realizations are used for low-density parity-check (LDPC) codes, for example.

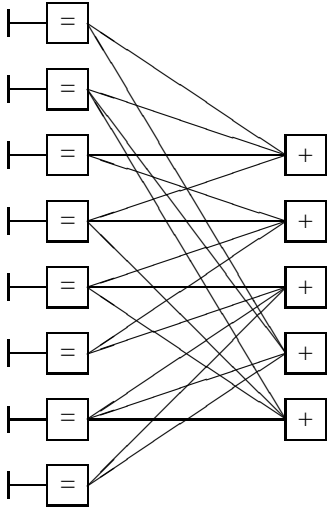
\begin{figure}[h]
\setlength{\unitlength}{5pt}
\centering
\begin{picture}(40,39)(30,0)
\multiput(42.5,0.5)(0,5){8}{\framebox(3,3){=}}
\multiput(42.5,2)(0,5){8}{\line(-1,0){3}}
\multiput(39.5,1)(0,5){8}{\line(0,1){2}}
\multiput(60.5,5.5)(0,5){5}{\framebox(3,3){+}}
\put(60.5,27){\line(-3,0){15}}
\put(60.5,27){\line(-3,-1){15}}
\put(60.5,27){\line(-3,1){15}}
\put(60.5,27){\line(-3,2){15}}
\put(60.5,22){\line(-3,1){15}}
\put(60.5,22){\line(-3,-1){15}}
\put(60.5,22){\line(-3,-2){15}}
\put(60.5,22){\line(-3,0){15}}
\put(60.5,17){\line(-3,-1){15}}
\put(60.5,17){\line(-3,0){15}}
\put(60.5,17){\line(-3,-2){15}}
\put(60.5,17){\line(-1,-1){15}}
\put(60.5,12){\line(-3,-2){15}}
\put(60.5,12){\line(-3,-1){15}}
\put(60.5,12){\line(-3,5){15}}
\put(60.5,12){\line(-3,4){15}}
\put(60.5,7){\line(-3,2){15}}
\put(60.5,7){\line(-3,0){15}}
\put(60.5,7){\line(-1,1){15}}
\put(60.5,7){\line(-3,5){15}}
\end{picture}

\caption{Uncontrollable parity-check realization of $(8,4,4)$ code.}
\label{RM}
\end{figure}

\vspace{1ex}
\noindent
\textbf{Example 2}.  The five binary 8-tuples $11110000$, $00111100$, $00001111$, $11000011$, $01011010$ form a set of five linearly dependent generators for  the $(8,4,4)$ first-order Reed-Muller (RM) code. 
Since $\CC$ is self-dual (\ie $\CC^\perp = \CC$), the same five binary 8-tuples form a set of linearly dependent checks for $\CC$.  The parity-check realization based on these five check 8-tuples is shown in Fig.\ \ref{RM}.  

\vspace{1ex} 
\noindent
\textbf{Theorem 5}.  A generator realization is controllable, and a parity-check realization is observable.  A generator realization is observable if and only if its generators are linearly independent, and a parity-check realization is controllable if and only if its checks are linearly independent.

\vspace{1ex}
\noindent
\textit{Proof}:  A parity-check realization is obviously observable, since all state variables are (multiples of) replicas of symbol variables (see, \eg Fig.\ \ref{RM}).  By observability/controllability duality, a generator realization must therefore be controllable.

In a generator realization, there is a nonzero configuration $(\zerob, \sb)$ if and only if there is some linear combination of the generators that equals the zero codeword $\zerob \in \CC$, which happens if and only if the generators are linearly dependent.  Thus a generator realization is observable if and only if its generators are linearly independent.  By observability/controllability duality, a parity-check realization is controllable if and only if its checks are linearly independent.  \qed \vspace{1ex}

For example, the parity-check realization of Fig.\ \ref{RM} is observable but uncontrollable.  So is the realization of Fig.\ \ref{Fig2a}(b), which may be viewed as a parity-check realization with the three redundant check 3-tuples $110, 011, 101$.

\subsection{Local reducibility}
We now show how to locally reduce any unobservable or uncontrollable realization.

\vspace{1ex}
\noindent
\textbf{Theorem 6}. An unobservable linear  realization with a nonzero trajectory $(\zerob, \sb) \in \Bf$ may be locally reduced by trimming any single state space in the support of $\sb$.  The dual uncontrollable realization may be correspondingly locally reduced by the dual merging operation. 

 \vspace{1ex}
 \noindent
 \textit{Proof}:  Select any state space $\SSS_j$ such that $s_j \neq 0$ in some unobservable trajectory $(\zerob, \sb) \in \Bf$.    Choose a basis $\{g_{j\ell} \}$ for $\SSS_j$ with $g_{j1} = s_j$.  The coordinates of $s_j$ are thus $10\ldots0$.  
 Define the subspace $\cT_j \subset \SSS_j$ as the set of all $s_j \in \SSS_j$ such that $s_{j1} = 0$, and trim the realization by restricting $\Bf$ to the subbehavior $\Bf'$ consisting of those trajectories that pass through a state $s_j \in \cT_j$.  We may then replace $\SSS_j$ by $\cT_j$, reducing the state space dimension by one.  
The trajectory $(\zerob, \sb)$ is then not in the trimmed behavior $\Bf'$, since $s_{j1} = 1$.
 However, given any $(\ab, \sb') \in \Bf$, the entire coset $\{(\ab, \sb' + \alpha \sb): \alpha \in \F\}$ is in $\Bf$ and thus realizes $\ab$.  In this coset, the value of  $S_{j1}$ runs through $\{s_{j1}' + \alpha s_{j1}:  \alpha \in \F\} = \F$, so precisely one trajectory $(\ab, \sb' + \alpha \sb)$ has first state coordinate $s_{j1}' + \alpha s_{j1} = 0$.  Thus the trimmed realization still realizes every $\ab \in \CC$.

In the dual realization, the corresponding local reduction is the merging of the states in $\hat{\SSS}_j$ to their cosets in $\hat{\SSS}_j/(\cT_j)^\perp$ via the natural map.  In other words, the  first coordinate of $\hat{\SSS}_j$ is simply deleted.  Since the trimmed primal realization generates $\CC$, the merged dual realization must still generate $\CC^\perp$.  \qed \vspace{1ex}

\subsection{Tail-biting trellis realizations}

Next, we consider linear tail-biting trellis realizations, which form the simplest class of realizations on a graph with a cycle.  We will see that the effects of uncontrollability in this case are similar to those seen in classical state-space (conventional trellis) realizations, namely:

\vspace{1ex}
\noindent
\textbf{Theorem 7}.  A reduced linear tail-biting trellis realization is uncontrollable if and only if its behavior consists of disconnected subbehaviors. 

 \vspace{1ex}
 \noindent
 \textit{Proof}:  The dual unobservable tail-biting trellis realization has a nonzero trajectory $(\zerob, \sb) \in \Bf$.
Moreover, $s_i \neq 0$ for all $i$, since otherwise some constraint code $\CC_i$ would not be proper.
We again choose a basis for each state space $\SSS_i$ with $g_{i1} = s_i$, so the coordinates of each $s_i$ are $10\ldots0$.
If we choose a dual basis for each state space $\hat{\SSS}_i$ in the dual uncontrollable realization,
then the inner product of elements of $\SSS_i$ and $\hat{\SSS}_i$ is given by the dot product of their coordinate vectors.  Thus the set $s_i^\perp \subseteq \hat{\SSS}_i$ of dual states orthogonal to $s_i$ is the subspace $\hat{\cT}_i$ of $\hat{\SSS}_i$ consisting of states whose first coordinate is zero.  The $|\F|$ cosets of $\hat{\cT}_i$ in $\hat{\SSS}_i$ are the subsets of $\hat{\SSS}_i$ whose first coordinates equal a certain value of $\F$.

Each constraint code $\CC_i$ thus contains an element $(s_i, a_i, s_{i+1})$ with $a_i = 0$ and $s_i$ and $s_{i+1}$ having coordinates $10\ldots0$.  It follows that if $(\hat{s}_i, \hat{a}_i, -\hat{s}_{i+1})$ is any element of the orthogonal code $\CC_i^\perp$ (using the convention that the sign inversion is always applied to $\hat{s}_{i+1}$), then
\begin{eqnarray*}
0 & = & \langle{(s_i, a_i, s_{i+1})},{(\hat{s}_i, \hat{a}_i, -\hat{s}_{i+1})}\rangle \\ & = & \langle{s_i},{\hat{s}_i}\rangle + \langle{a_i},{\hat{a}_i}\rangle - \langle{s_{i+1}},{\hat{s}_{i+1}}\rangle \\ & = & \hat{s}_{i1} - \hat{s}_{i+1,1}.
\end{eqnarray*}
 It follows that in any dual trajectory $(\hat{\ab}, \hat{\sb})$, the first coordinates of the dual state variables must be equal: $\hat{s}_{i1} = \hat{s}_{i+1,1}$.
 Thus the state spaces $\hat{\SSS}_i$ are partitioned by their first coordinates into $|\F|$ cosets of $\hat{\cT}_i$, such that state transitions are possible only within cosets.  The trajectories in the dual behavior $\Bf^\circ$ thus partition into disconnected cosets, each consisting of the trajectories that go through states with a particular value of the first coordinate.  \qed

\begin{figure}[h]
\setlength{\unitlength}{5pt}
\centering
\begin{picture}(40,17)(-7, -1)
\put(0,0){\circle*{1}}
\put(-4,-0.5){00}
\put(-4,4.5){01}
\put(-4,9.5){10}
\put(-4,14.5){11}
\put(0,5){\circle*{1}}
\put(0,10){\circle*{1}}
\put(0,15){\circle*{1}}
\put(0,0){\line(1,0){5}}
\put(0,5){\line(1,-1){5}}
\put(2,0){0}
\put(2,3){1}
\put(0,10){\line(1,0){5}}
\put(0,15){\line(1,-1){5}}
\put(2,10){0}
\put(2,13){1}
\put(5,0){\circle*{1}}
\put(5,10){\circle*{1}}
\put(5,10){\line(1,0){5}}
\put(5,0){\line(1,0){5}}
\put(5,10){\line(1,-2){5}}
\put(5,0){\line(1,2){5}}
\put(7,0){0}
\put(7,10){0}
\put(6,6){1}
\put(8,6){1}
\put(10,0){\circle*{1}}
\put(10,10){\circle*{1}}
\put(10,0){\line(1,0){5}}
\put(10,10){\line(1,0){5}}
\put(10,0){\line(1,1){5}}
\put(10,10){\line(1,1){5}}
\put(12,0){0}
\put(12,3){1}
\put(12,10){0}
\put(12,13){1}
\put(15,10){\circle*{1}}
\put(15,15){\circle*{1}}
\put(15,0){\circle*{1}}
\put(15,5){\circle*{1}}
\put(15,0){\line(1,0){5}}
\put(15,5){\line(1,1){5}}
\put(15,0){\line(1,1){5}}
\put(15,5){\line(1,2){5}}
\put(17,0){0}
\put(17,3){1}
\put(19,8){1}
\put(19,12){0}
\put(15,10){\line(1,0){5}}
\put(15,15){\line(1,-3){5}}
\put(15,10){\line(1,1){5}}
\put(15,15){\line(1,-2){5}}
\put(19,10){0}
\put(17,13){1}
\put(19,6){0}
\put(19,1){1}
\put(20,10){\circle*{1}}
\put(20,15){\circle*{1}}
\put(20,0){\circle*{1}}
\put(20,5){\circle*{1}}
\put(20,0){\line(1,0){5}}
\put(20,5){\line(1,0){5}}
\put(20,10){\line(1,0){5}}
\put(20,15){\line(1,0){5}}
\put(22,0){0}
\put(22,5){0}
\put(22,10){0}
\put(22,15){0}
\put(20,0){\line(1,1){5}}
\put(20,5){\line(1,-1){5}}
\put(20,10){\line(1,1){5}}
\put(20,15){\line(1,-1){5}}
\put(21,3){1}
\put(23,3){1}
\put(21,13){1}
\put(23,13){1}
\put(25,10){\circle*{1}}
\put(25,15){\circle*{1}}
\put(25,0){\circle*{1}}
\put(25,5){\circle*{1}}
\end{picture}
\caption{Unobservable linear tail-biting trellis realization. }
\label{FigU}
\end{figure}
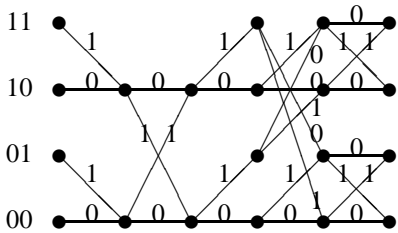

\vspace{1ex}
\noindent
\textbf{Example 3}.  The tail-biting trellis realization of Fig.\ \ref{FigU} realizes the binary linear $(5,3)$ block code $\CC = \langle 01110$,  $10010$, $01101 \rangle$.    Since it has a nonzero trajectory $(\zerob, \sb)$, this realization is unobservable.  Its five state spaces have been coordinatized so that the state values along the nonzero trajectory $(\zerob, \sb)$ are either 10 or 1. 

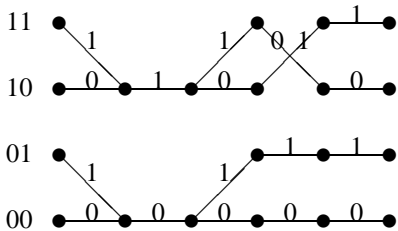
\begin{figure}[h]
\setlength{\unitlength}{5pt}
\centering
\begin{picture}(40,17)(33, -1)
\put(36,-0.5){00}
\put(36,4.5){01}
\put(36,9.5){10}
\put(36,14.5){11}
\put(40,0){\circle*{1}}
\put(40,5){\circle*{1}}
\put(40,10){\circle*{1}}
\put(40,15){\circle*{1}}
\put(40,0){\line(1,0){5}}
\put(40,5){\line(1,-1){5}}
\put(42,0){0}
\put(42,3){1}
\put(40,10){\line(1,0){5}}
\put(40,15){\line(1,-1){5}}
\put(42,10){0}
\put(42,13){1}
\put(45,0){\circle*{1}}
\put(45,10){\circle*{1}}
\put(45,10){\line(1,0){5}}
\put(45,0){\line(1,0){5}}
\put(47,0){0}
\put(47,10){1}
\put(50,0){\circle*{1}}
\put(50,10){\circle*{1}}
\put(50,0){\line(1,1){5}}
\put(50,10){\line(1,1){5}}
\put(50,0){\line(1,0){5}}
\put(50,10){\line(1,0){5}}
\put(52,0){0}
\put(52,3){1}
\put(52,10){0}
\put(52,13){1}
\put(55,10){\circle*{1}}
\put(55,15){\circle*{1}}
\put(55,0){\circle*{1}}
\put(55,5){\circle*{1}}
\put(55,0){\line(1,0){5}}
\put(55,5){\line(1,0){5}}
\put(55,15){\line(1,-1){5}}
\put(55,10){\line(1,1){5}}
\put(57,0){0}
\put(57,5){1}
\put(56,13){0}
\put(58,13){1}
\put(60,10){\circle*{1}}
\put(60,15){\circle*{1}}
\put(60,0){\circle*{1}}
\put(60,5){\circle*{1}}
\put(60,0){\line(1,0){5}}
\put(60,10){\line(1,0){5}}
\put(62,0){0}
\put(62,10){0}
\put(60,5){\line(1,0){5}}
\put(60,15){\line(1,0){5}}
\put(62,15){1}
\put(62,5){1}
\put(65,10){\circle*{1}}
\put(65,15){\circle*{1}}
\put(65,0){\circle*{1}}
\put(65,5){\circle*{1}}
\end{picture}
\caption{Dual uncontrollable linear tail-biting trellis realization. }
\label{FigUb}
\end{figure}

The dual realization of Fig.\ \ref{FigUb} realizes the orthogonal $(5,2)$ code $\CC^\perp = \langle 10111, 01100 \rangle$.  This is a product realization with generators 
$\langle \underline{1}0\underline{111}$, $\underline{01100}\rangle$, where the span of the second generator is \emph{degenerate} (\ie all states are nonzero).  It has two  disjoint subbehaviors, as predicted by Theorem 7.  \qed \vspace{1ex}

The fundamental structure theorem of Koetter and Vardy \cite{KV03} states that every linear reduced tail-biting (or conventional) trellis realization is a product realization.  In this light, it is easy to see that a tail-biting trellis is controllable if and only if none of the spans of its generators is degenerate.

\subsection{General realizations}

We have seen that for trellis realizations, the effects of uncontrollability for conventional and tail-biting trellises are similar.  However, for more general realizations, we do not necessarily have the same phenomenon of disconnected subbehaviors.  For example, in the uncontrollable parity-check realization of Fig.\ \ref{RM}, every state variable is a replica of a symbol variable, and thus (from the properties of $\CC$) every pair of state variables that do not both correspond to the same symbol variable can take on all possible pairs of values.  

Unobservable realizations would appear to be clearly undesirable for use with iterative decoding:  because every code sequence has multiple representations, it seems that an iterative decoding algorithm would never converge to any one of them.  However, we have seen that any unobservable realization may be locally reduced, so without loss of generality we can always assume an observable realization for any given code.

On the other hand, uncontrollable parity-check realizations can be and have been used successfully with iterative decoding.  Moreover, it is known that redundant parity checks reduce the number of pseudocodewords with iterative decoding of LDPC codes, so in this respect uncontrollability may actually be helpful.  Thus it would be useful to have a better understanding of the properties of uncontrollable realizations.

\end{document}